\documentclass{ws-p8-50x6-00}
 
\DeclareMathAlphabet   {\mathsc}{OT1}{cmr}{m}{sc} 
 
\def\[{\left [} 
\def\]{\right ]} 
\def\({\left (} 
\def\){\right )} 
\newcommand{\lang}{\left\langle} 
\newcommand{\rang}{\right\rangle} 
\newcommand{\lbr}{\left\{} 
\newcommand{\rbr}{\right\}} 
\newcommand{\oline}[1]{\overline{#1}}

\newcommand{\wtd}[1]{\widetilde{#1}} 

\newcommand{\GS}       {\mathsc{gs}} 
 
\newcommand{\PV}       {\mathsc{pv}}

\newcommand{\hc}       {\mathrm{\; h.c. \;}}

\newcommand{\re}{{\rm Re}}

\newcommand{\gappeq}{\mathrel{\rlap {\raise.5ex\hbox{$>$}} 
{\lower.5ex\hbox{$\sim$}}}} 
\newcommand{\lappeq}{\mathrel{\rlap{\raise.5ex\hbox{$<$}} 
{\lower.5ex\hbox{$\sim$}}}}

\def\Eisen{G_{2}\(t,\bar{t}\)}

\renewcommand{\thepage}{\arabic{page}} 
\def\Eisen{G_{2}\(t,\bar{t}\)} 

\begin{document}

\title{Phenomenological Aspects of Heterotic Effective Models at One Loop}


\author{\underline{Y. Mambrini}, P. Bin\'etruy, A. Birkedal Hansen and
B. D. Nelson}

\address{Departamento de F\'isica Te\'orica C-XI, Universidad Aut\'onoma de Madrid, Cantoblanco, 28049 Madrid, Spain.
\\E-mail: mambrini@delta.ft.uam.es}


\maketitle

\abstracts{We provide a study of the phenomenology of orbifold
compactifications of the heterotic string within the context of
supergravity effective theories. Our investigation focuses on
those models where the soft Lagrangian is dominated by loop
contributions to the various soft supersymmetry breaking
parameters. We consider the pattern
of masses that are governed by these soft terms and investigate
the implications of certain indirect constraints on supersymmetric
models, such as flavor-changing neutral currents, the anomalous
magnetic moment of the muon and the density of thermal relic
neutralinos.}

\section{Introduction}

One of the most crucial and difficult tasks of string phenomenologists is now to 
make, and keep, contact between the high energy theory, and the low energy world. 
For that, we need to consider a superstring theory which yields in four dimensions, 
the Standard Model gauge group,  three generations of quarks, and a consistent 
mechanism of SUSY breaking. Our analysis have relies on 
orbifold compactifications of the heterotic string within the context of 
supergravity effective theories. More specifically, we concentrate on those 
models where the 
action is dominated by one loop order contributions to soft breaking terms. 
Recently, all one loop order contributions have been calculated 
\cite{BiGaNe01}. The key point of such models is the non universality of
supersymmetry breaking term which is a  
consequence of the beta--function appearing in the superconformal anomalies. 
This non universality gives a specific phenomenology in the gaugino 
and the scalar sectors, modifying the predictions coming from Msugra. 
In fact, these string--motivated models show new behavior that interpolates 
between the phenomenology of unified supergravity models (Msugra) and models 
dominated by the superconformal anomalies (AMSB). The constraints arising from 
accelerator physics, and dark matter aspects have been already studied \cite{Bin1}.
 It becomes interesting now, to see in which sense 
experimental limits on supersymmetric particles  will be able to
 bring us informations, or even to rule out some of these models.

\section{Theoretical Framework}

Our phenomenological study is based on orbifold compactifications of the 
weakly--coupled heterotic string, where we distinguish two regimes.
In the first one, SUSY breaking is driven by the compactification
 moduli $T$, whose vacuum expectation values determine the size of the 
 compact manifold. In the second one, it is the dilaton field $S$, whose vacuum 
 expectation value determines the magnitude of the (unified) coupling constant 
 $g_{\mathrm{STR}}$ at the string scale, that transmits, via its auxiliary fields,
  SUSY breaking. We work in the context of models in which 
  string nonperturbative corrections to the Kahler potential act to stabilize the
   dilaton in the presence of gaugino condensation
\cite{BiGaWu96,BiGaWu97a}. The origins of breaking terms are diverse. 
Some coming from the superconformal anomalies are non--universal (proportional to
 the
 beta-- function of the $SU(3) \times SU(2) \times U(1)$ groups) some are 
 independent of the gauge group considered (Green--Schwarz counterterm, $vev$ 
 of the condensate). This interplay between universality and non--universality 
 gives a rich new phenomenology, and indicates new trends in the
  search of supersymmetric particles in accelerator or astroparticle physics. 
\subsection{The moduli dominated scenario}
In the moduli dominated scenario, the supersymmetric susy breaking terms can 
be written\cite{BiGaNe01,GaNeWu99,GaNe00b}
\begin{eqnarray}
M_a &=& \frac{g_{a}^{2}\(\mu\)}{2} \lbr 2
 \[ \frac{\delta_{\GS}}{16\pi^{2}} + b_{a}
\]\Eisen F^{T} + \frac{2}{3}b_{a}\oline{M} \rbr, \label{modsoftgaugi}\\
A_{ijk}&=& - \frac{1}{3} \gamma_{i}\oline{M} - p \gamma_{i} \Eisen
F^{T} + {\rm cyclic}(ijk), \\ M_{i}^{2} &=& (1-p)\gamma_i
\frac{|M|^2}{9}. \label{modsoftscal}
\end{eqnarray}

where $b_a$ is the one loop beta--function coefficient of the
$SU(3)\times SU(2)\times U(1)$ gauge coupling $g_{a=1,2,3}$. 
The field $M$ is the auxiliary field of the supergravity multiplet 
related to the gravitino mass by

\begin{equation}
M_{3/2}=-\frac{1}{3}\langle \overline{M} \rangle.
\end{equation}

We clearly see in these formulae the mixing between universal term 
and non--universal ones. Moreover, scalar mass terms are coming with a 
loop suppression factor $\gamma_i$, and the gaugino mass breaking terms 
have a
 universal compensation coming from the Green--Schwarz counterterm 
 (appearing in order to cancel anomalies) that 
 can give high value to the chargino or neutralino masses. To sum up, this 
 regime gives light scalars and relatively heavy gauginos, whose nature 
 depends completely on the value of $\delta_{\mathrm{GS}}$.

\subsection{The dilaton dominated scenario}

In this region of parameter space, we can express the soft SUSY breaking
terms as

\begin{eqnarray}
M_{a}&=&\frac{g_{a}^{2}\(\mu\)}{2} \lbr  \frac{2}{3}b_{a}\oline{M}
+\[ 1 - 2 b_{a}' k_s \] F^{S} \rbr \label{dilatsoftgaugi}\\ A_{ijk} &=&
-\frac{k_s}{3}F^S - \frac{1}{3} \gamma_{i}\oline{M} +
\tilde{\gamma}_{i} F^{S} \lbr \ln(\mu_{\PV}^{2}/\mu_R^2)
-p\ln\[(t+\bar{t}) |\eta(t)|^4\] \rbr + (ijk) 
 \\ M_{i}^{2} &=& \frac{|M|^2}{9}
 \[ 1 + \gamma_i
-\(\sum_{a}\gamma_{i}^{a} -2\sum_{jk}\gamma_{i}^{jk}\) \(
\ln(\mu_{\PV}^{2}/\mu_R^2) -p\ln\[(t+\bar{t}) |\eta(t)|^4\] \) \]
\nonumber
 \\
 & &+ \lbr
 \wtd{\gamma}_{i}\frac{MF^S}{6}+\hc \rbr , \label{dilatsoftscal}
\end{eqnarray}

\noindent
with

\begin{equation}
F^{S} = 3
\frac{\frac{2}{3}b_{+}}{1-\frac{2}{3}b_{+}K_{s}} M_{3/2}.
\label{FS}
\end{equation}

\noindent
with $b_+$ being the largest beta--function coefficient among the condensing
gauge groups of the hidden sector, $k_s$ the derivative in $S$ 
of the Khaler potential and $p_i$ the Pauli--Villars weigths of the regulator fields.

The phenomenology of the dilaton dominated scenario is completly different
from the moduli dominated one. If we look at (\ref{dilatsoftscal}) and 
(\ref{dilatsoftgaugi}),
 it
is clear that we are in a domain of heavy squarks and sleptons (of the order
of the gravitino scale) and light gaugino masses, directed by the dilaton
auxiliary field $vev's$. Indeed, the beta--functions $b_a$ are of the order 
of $10^{-2}$,
which will not be competitive compared to the $F$ term of the dilaton in
(\ref{dilatsoftgaugi}). In fact, if we look more clearly at
(\ref{FS}), for not so big values of $b_+$, we can consider that $F^S$ has a
 linear evolution as a function of $b_+$. Increasing $b_+$ means approaching
 the universal case for the gaugino sector (and the scalar one, driven by 
 $M_{3/2}$).

\begin{figure}
\epsfxsize=27pc 
\epsfbox{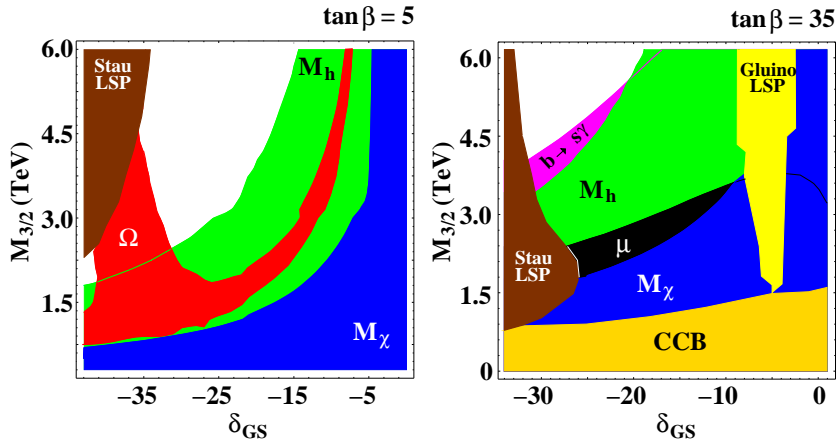}
\caption{{\bf Constraints on the moduli-dominated parameter space for $\tan\beta=5$
 (left) and $\tan\beta=35$ (right) with $p = 0$ and 
$\lang \re \; t \rang = 2.0$}. Constraints on the ($M_{3/2},\; \delta_{\GS}$)
 plane are given for $\mu > 0$. The dark shaded regions on the left have a
  stau LSP. For $\tan\beta=5$ the region labeled ``$\Omega$'' has the
  cosmologically preferred relic density of neutralinos. The exclusion 
  contours are due to (from bottom right to upper left) CCB vacua, the
   chargino mass, too large SUSY contributions to $(g_{\mu}-2)$, the Higgs
    mass limit and too large a $b \to s \gamma$ rate. }\label{fig:fig1}
\end{figure}
\begin{figure}
\epsfxsize=27pc  
\epsfbox{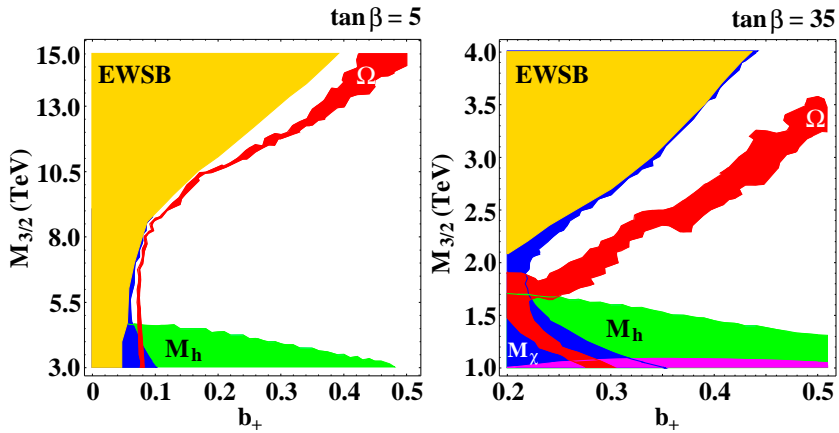}
\caption{ {\bf Constraints on the dilaton-dominated parameter space for
 $\tan\beta=5$ (left) and $\tan\beta=35$ (right).} Constraints on the
($M_{3/2},\; b_+$) plane are given for $\mu > 0$. \label{fig:fig2}}
\end{figure}

Even within the restricted context of the weakly-coupled heterotic
string compactified on an orbifold, the phenomenology of such
supergravity effective theories is far richer and varied than that
of the standard ``minimal'' supergravity approach. While our
current indirect knowledge on the nature of supersymmetry already
constrains these models significantly, there still exist large
regions of parameter space for all the cases studied here -- even
for low values of $\tan\beta$. We anticipate that future
measurements of, or limits on, superpartner masses and kinematic
distributions will constrain the parameters of these models
further. So will future cosmological or astrophysical
measurements. In fact, this type of measurement forms a useful
complement that will be crucial in unraveling the nature of
supersymmetry breaking and its mediation in such models. 

Parameters that are related to the orbifold can have a large
impact on these broad features. This is a welcome result --
implying that experimental data can indeed probe the nature of the
underlying theory within a class of models. For example, in cases
where supersymmetry breaking is transmitted predominantly by the
moduli associated with compactification the relative sign of terms
proportional to their auxiliary field and that of the conformal
anomaly can in principle be measured. This parameter, in turn, is
related to the spontaneous breakdown of modular invariance that
may occur in such models.

Future studies of such string-based models should
focus on extending this initial survey to true collider signatures
in both hadron and lepton machines \cite{NelsonTeva}, 
as well as computing event rates for astrophysical processes \cite{Relicb}.

I would like to thank P. Bin\'etruy for its corrections and helpful discussions. 
This work would not have been possible without the constant help of G. Belanger.


\begin{thebibliography}{99}
\bibitem{BiGaNe01} 
  {\rm P.~Binetruy, M.~K.~Gaillard and B.~D.~Nelson}, 
  {\it Nucl. Phys.} {\bf B604} {\rm (2001) 32}. 
\bibitem{Bin1}
{\rm P.~Binetruy, A.~Birkedal-Hansen, Y.~Mambrini, B.D.~Nelson}
{arXiv:hep-ph/0308047}.
\bibitem{BiGaWu96} 
  {\rm P.~Bin\'{e}truy, M.~K.~Gaillard and Y.-Y.~Wu}, 
  {\it Nucl. Phys.} {\bf B481} {\rm (1996) 109}. 
\bibitem{BiGaWu97a} 
  {\rm P.~Bin\'{e}truy, M.~K.~Gaillard and Y.-Y.~Wu}, 
  {\it Nucl. Phys.} {\bf B493} {\rm (1997) 27}. 
\bibitem{GaNeWu99} 
  {\rm M.~K.~Gaillard, B.~D.~Nelson and Y.-Y.~Wu}, 
  {\it Phys. Lett.} {\bf B459} {\rm (1999) 549}. 
\bibitem{GaNe00b} 
  {\rm M.~K.~Gaillard and B.~D.~Nelson}, 
  {\it Nucl. Phys.} {\bf B588} {\rm (2000) 197}.
\bibitem{NelsonTeva}
{\rm G.L.~Kane, J.~Lykken, S.~Mrenna, B.D.~Nelson, L.T.~Wang, T.T.~Wang}
{\it Phys.Rev.} {\bf D67} {\rm (2003) 045008} 
{arXiv:hep-ph/0209061}.
\bibitem{Relicb} {\rm P.~Bin\'etruy, Y.~Mambrini, E.~Nezri}
{arXiv:hep-ph/0312155}.
\end{thebibliography}
\end{document}